# Mining the Air -- for Research in Social Science and Networking Measurement

**Scott Kirkpatrick, Hebrew University**   *kirk@cs.huji.ac.il*
**Ron Bekkerman and Adi Zmirli, Haifa University**   *ron.bekkerman@gmail.com*
**Francesco Malandrino, Politecnico de Torino**   *francesco.malandrino@polito.it*

 Smartphone apps provide a vitally important opportunity for monitoring human mobility, human experience of ubiquitous information aids, and human activity in our increasingly well-instrumented spaces.  As wireless data capabilities move steadily up in performance, from 2&3G to 4G (today's LTE) and 5G, it has become more important to measure human activity in this connected world from the phones themselves.  The newer protocols serve larger areas than ever before and a wider range of data, not just voice calls, so only the phone can accurately measure its location.  Access to the application activity permits not only monitoring the performance and spatial coverage with which the users are served, but as a crowd-sourced, unbiased background source of input on all these subjects, becomes a uniquely valuable resource for input to social science and government as well as telecom providers.

 The public also stands to benefit.  National and regional regulators tasked to ensure that consumers are getting the communications bandwidth, coverage and capability that were advertised and they paid for, are beginning to use crowd-sourced measurements from the edge to provide public "report cards" of communications quality.[1] We have been working with data captured by applications based on the phones, authorized by their users to capture location information and share it to build a public database of internet access performance.  We have used most extensively results from an Israeli startup called WeFi[2], which observes the category of application in use during some of its measurements and determines upload and download data volumes and rates.  We have, in all, about 3 billion measurements from five US cities and their surroundings,  Atlanta, Boston, Brooklyn, Los Angeles and San Francisco, for several months in each location during 2014 and 2015.

Our data has been presented in several publications that address issues in mobile network planning and management.[3,4]  One surprising result is the range over which each cell antenna is received.  Earlier studies in which the data source is a carrier have used the cell tower locations as a proxy for user location.[5,6]  In this study, which sees all carriers in each city, we first estimated the locations of the cell towers, which were named in each measurement record, as the centroid of all the observations where a tower was seen.  Cell dimensions of several km are observed.   By contrast, the phone locations are known to GPS accuracy, with errors as little as 10m.

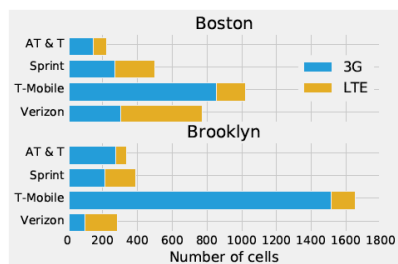 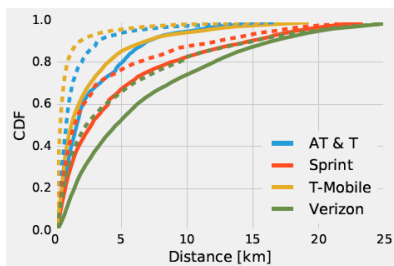 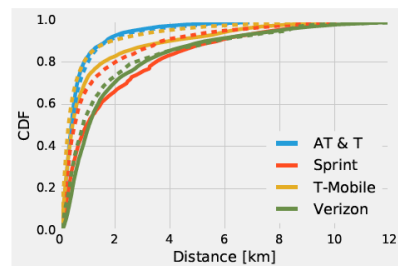

(a)   (b)   (c)

Fig. 1.  Number of LTE and 3G cells deployed by each operator (a); distance from user to cell tower in Boston (b) and Brooklyn (c).  Solid lines are LTE connections, dashed lines are 3G.

In this report we use unbiased observations made in the background in the course of the daily lives of over a hundred thousand people in and around Los Angeles, sampling roughly 1 per cent of the population, from all economic levels and demographics.  The WeFi application monitors location, data connections, and application usage on Android smartphones, but does not capture any content exchanged or any information relating to phone calls. Personally identifying information is removed from the data by hashing the identifiers of the phones.  We observe the activities of individual phones, but collect them into aggregated communities sufficient in number to prevent re-identification of individuals.  One purpose of this study is to understand how much data is required in order to observe the social behaviors relevant to well-functioning cities.  One observation is that the more data we can consider, the finer the scale which we can study without danger of compromising individuals' privacy.

Measurements based on the cellphone can occur whenever there is activity, either data transfer or motion of the user, and thus are much more frequent than the monitoring normally seen in datasets which record the metadata of phone calls.  In mobile CDRs and related carrier data the towers see each phone typically 5-20 times a day, with phones in cars seen more often as they change towers during a call.  Some of the WeFi data is taken much more frequently.  The Android operating system allows the WeFi app to request a measurement when the phone position changes by as little as .0001 in the Lat or Lon coordinate, or on any change in the system's connectivity.  As a result we are presented with position information as often as more than a thousand times an hour.

**Using Location Data**

In employing this data for more traditional social science ends we follow a methodology of successive reduction to isolate distinctive communities, then extract their characteristic patterns of commuting, working, shopping, and leisure activities. Our filters are simple and fairly strong. We have used two data sets.  In the first, we observe over 131K users, recording their position as lat/lon to a precision of .0001, with a timestamp giving days, minutes and seconds.  The 835M measurements in this data set take up about 20 GB.  We refer to this data as the location data set.  It was gathered during March through May 2015,.  The second data set, collected in February, 2015 consisted of 130 GB with 422M measurements, each containing location information plus details of the applications in use, the data connection used, and the amounts of data uploaded or downloaded since the previous measurement.  We refer to this data as the application data set.  Because these measurements were grouped an hour at a time to simplify their retrieval, the time stamps were only given to the precision of an hour.  The application data set contained over 91K distinct users.  Each user is only known to us by a random hash of the machine identification of their Android phone. The same hash function was used in gathering the two data sets, so we can determine that over 67K of the UIDs (the user's randomized identifier) are present in both data sets.  This allows us to combine information gathered in each of these two ways. We start our characterization of interesting communities with the location data set.We refine the 57K UIDs by determining the towns or neighborhoods in which they reside.  The criterion applied was that a residence is identified when the UID is seen at the same location to an accuracy of .001 in lat/lon two or more hours a night for 15 or more nights within our sample.  For more than 51K of our consistent UIDs we can identify such home locations.  More than half of the UIDs ( >36K) live outside the city of LA proper, and are called suburbanites in this discussion.  However, the city of LA is rather porous. Thus quite a few of our suburban districts are governed as separate towns but are contained within the city of LA.  More than 10K UIDs reside in neighborhoods in the city proper.  We also omitted some UIDs for which the suburb or neighborhood is not unambiguously defined, or for which the residential area population is <5000 people.  This left us with 36,531 users who live in 232 independent towns and 10,573 users who live in 83 different neighborhoods within the city.

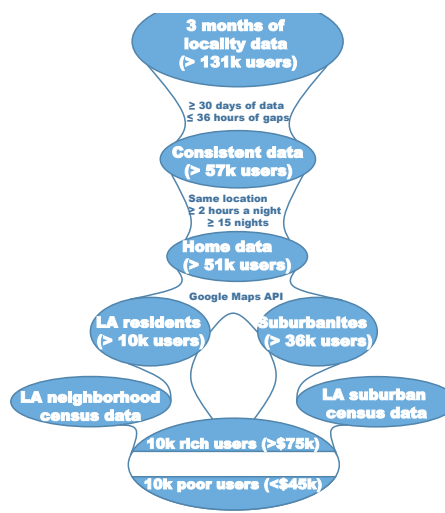

Next, using census demographic tables of population and median income, we assign to each UID the median income of their home district.  A concern is that usage of smart phones might skew our sample of users within the populations of each home town, but with smart phone usage now passing 60-70% of all mobile telephone customers, we do not think this will cause significant bias. We next identify a population of 10,094 UIDs (and their users) from the bottom of the demographics, living in areas with median income < $45K, and another 9780 UIDs, whose users live in the wealthiest areas, with median incomes > $75K.  We will distinguish these "rich" and "poor" users when identifying further details of their activities. The two sets make up 20.8% (the "poor" cohort) and 21.5% (the "rich" cohort) of our total sample of users.

We next determine where our users work. We again start with the users seen more than 30 days, with few gaps, the consistent user set.  We identify a stable daytime location, or work location, as a place defined to .001 accuracy in lat and lon, and seen for at least 4 fours per day, on at least 30 workdays. For almost 25K users, we can find such locations, but for 14K users, this was also their home location.  This leaves us with 10,596 commuters, for whom we also know their home location.  We still have 21.4% of our sample of users to study.  Of these, 2263 (or 20.9% of the commuters) live in our wealthier districts, thus are members of the rich cohort, and 1902 (0r 17.5% of the commuters) live in the poorer neighborhoods.  We notice that the second group has dropped by 4% or about one fifth, in their participation, as these members of the poorest cohort do not have fixed locations in which they work during the day.  The distribution of hours worked also shows an important difference.  Our richer cohort works shorter hours than the poor cohort.

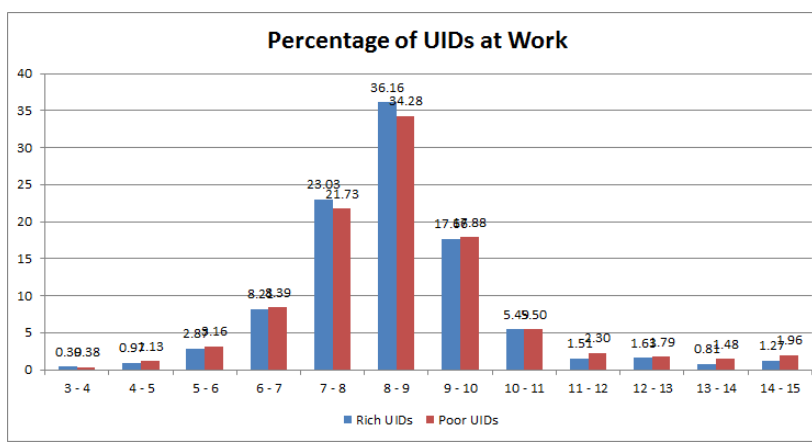

**Adding Application Data**

To find larger populations and address larger distinctions in behavior, we looked into two things that affect almost everyone: shopping and fast food restaurants. This required some manual effort. Shopping malls were identified with the help of Google Maps API, and resolved into 418 shopping areas. Over 37,000 of our consistent users spent at least 10 minutes and up to 6 hours at one of our shopping areas during the study period. They averaged 1 hour and 7 minutes per visit. Similarly, MacDonald's restaurants were screened to identify 553 with outdoor, separated locations, not inside some larger shopping center. McDonald's visitors spent from 5 minutes to two hours, for an average of 24 minutes in their fast food breaks. In order to know more about who goes to these, when, and why, we need some information about the users' interests. Here the Application data set is useful.

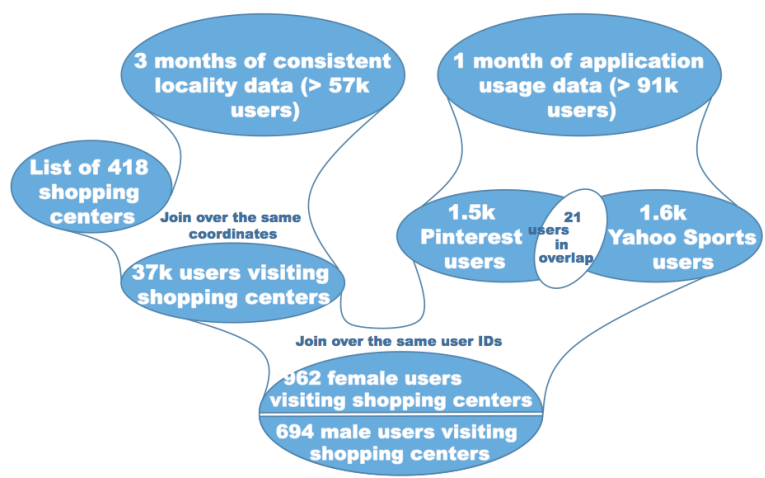

An indicator that serves to differentiate populations are the differences between two applications, Pinterest and Yahoo Sports. Pinterest has been found elsewhere to attract about 85% female users. Frequenting Yahoo Sports for results would seem to give a high probability that the user is a male. Almost all of the users we can separate in this way are seen at some point in our data set, shopping at a mall. The Fig at right summarizes the sizes of the communities that we extract in this fashion. We see 1.5K Pinterest users (each of whom has invoked the app >100 times) and 1.6K Yahoo sports users (also calling for the latest results at least 100 times). Many of these users, about 1000 of the likely female UIDs and almost 700 of the sports fans were seen at one of the 418 shopping centers.

Looking at the activity patterns, we see that our female shoppers are seen at the malls almost three times per week, while the males appear less than twice per week. In the longer work from which these examples are drawn we also separate younger and older users, analyze fast food consumption, commuting times and distances.

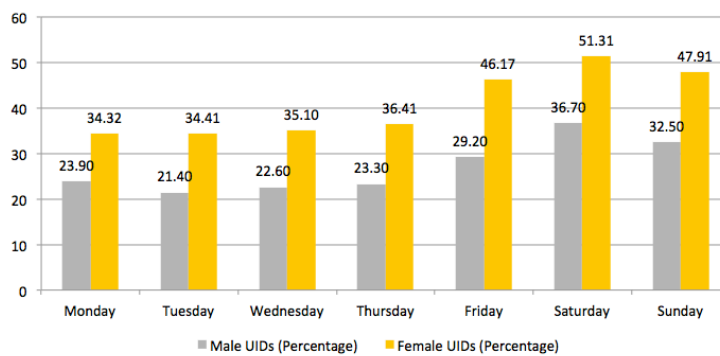

We have been able to resolve our data down by two and even three levels of filtering, but only because we have a lot of it. Imagine how much could be safely learned if all phones contributed to this information!

**Acknowledgements**

We are indebted to WeFi for the data on which this study was performed, and to Alex Zaidelson of WeFi for several productive discussions.  SK and FM were supported by the HUJI Cyber Security Center in conjunction with the Israel National Cyber Bureau in the Prime Ministers' Office.  Work by RB and AZ was supported by the Magnet Infomedia Grant of the Israel Ministry of Economy.